# On the Automated Detection of Corneal Edema with Second Harmonic Generation Microscopy and Deep Learning

Stefan R. Anton[1,2], Rosa M. Martínez-Ojeda[3], Radu Hristu[1], George A. Stanciu[1], Antonela Toma[2], Cosmin K. Banica[4], Enrique J. Fernández[3], Mikko Huttunen[5], Juan M. Bueno[3,*], Stefan G. Stanciu[1,*]

[1]Center for Microscopy-Microanalysis and Information Processing, University Politehnica Bucharest, 060042 Bucharest, Romania
[2]Center for Research and Training in Innovative Techniques of Applied Mathematics in Engineering, University Politehnica of Bucharest, 060042 Bucharest, Romania
[3]Laboratorio de Óptica, Campus de Espinardo (ed. 34) Universidad de Murcia, 30100 Murcia, Spain
[4]Faculty of Electrical Engineering, University Politehnica of Bucharest, 060042 Bucharest, Romania
[5]Photonics Laboratory, Physics Unit, Tampere University, Tampere, Finland

*E-mail addresses: bueno@um.es (J.M. Bueno), stefan.g.stanciu@upb.ro (S.G. Stanciu).

**Abstract:**

When the cornea becomes hydrated above its physiologic level it begins to significantly scatter light, loosing transparency and thus impairing eyesight. This condition, known as corneal edema, can be associated with different causes, such as corneal scarring, corneal infection, corneal inflammation, and others, making it difficult to diagnose and quantify. Previous works have shown that Second Harmonic Generation Microscopy (SHG) represents a valuable non-linear optical imaging tool to non-invasively identify and monitor changes in the collagen architecture of the cornea, potentially playing a pivotal role in future in-vivo cornea diagnostic methods. However, the interpretation of SHG data can pose significant problems when transferring such approaches to clinical settings, given the low availability of public data sets, and training resources. In this work we explore the use of three Deep Learning models, the highly popular InceptionV3 and ResNet50, alongside FLIMBA, a custom developed architecture, requiring no pre-training, to automatically detect corneal edema in SHG images of porcine cornea. We discuss and evaluate data augmentation strategies tuned to the specifics of the herein addressed application and observe that Deep Learning models building on different architectures provide complementary results. Importantly, we observe that the combined use of such complementary models boosts the overall classification performance in the case of differentiating edematous and healthy corneal tissues, up to an AU-ROC=0.98. These results have potential to be extrapolated to other diagnostics scenarios, such as differentiation of corneal edema in different stages, automated extraction of hydration level of cornea, or automated identification of corneal edema causes, and thus pave the way for novel methods for cornea diagnostics with Deep-Learning assisted non-linear optical imaging.

## 1. Introduction

The cornea represents the refractive front surface of the ocular globe. It's five layers, the epithelium, the Bowman's layer, the stroma, the Descemet's membrane and the endothelium have complementary functions[1], and the compromised integrity of any of these can lead to debilitating consequences up to complete vision loss. The corneal stroma is normally about 78% hydrated. When the cornea becomes hydrated with >5% above this physiologic level it begins to significantly scatter light, loosing transparency and thus impairing eyesight. Upon excessive hydration, the cornea swells only in a plane perpendicular to the surface, as swelling in the plane of the cornea is restricted by the collagen fibrils running in this direction[2]. This condition is known as corneal edema, and can originate from different causes, such as corneal scarring, corneal infection, corneal inflammation, corneal dystrophies, effects of administered drugs[3, 4] or surgical interventions[5], and others, making corneal edema difficult to diagnose and quantify. The impact of edema on patient comfort and eye function is devastating. Failing to diagnose it, and to proceed to timely treatment, can result in irreversible damage to the corneal ultrastructure, such as anterior corneal fibrosis or endothelial cell damage[6], which can lead to temporary or permanent vision impairment.

Second Harmonic Generation Microscopy is emerging as a powerful non-linear optical imaging tool to characterize the eye[7-14], holding significant promise to enable a new generation of diagnostic protocols that are more sensitive than current ones. Its huge potential for characterizing various parts of the eye, and of other tissues in the human body, originates from the ability to image in a label-free manner non-centrosymmetric structures, which exhibit a non-vanishing second-order susceptibility tensor $\chi^{(2)}$. When irradiated by ultrashort laser pulses, such structures generate a nonlinear optical signal at exactly half the wavelength of the excitation source[15]. Collagen represent the most abundant protein in the human body and the main structural protein in the extracellular matrix of animal tissues[16, 17], and exhibits a non-centrosymmetric structure yielding strong SHG signals that have been explored to date in a vast number of applications addressing various parts of the human body, such as epithelial tissues[18-21], muscles[22-24], bones[25-27], brain[28-30], etc. SHG's potential to characterize the collagen in tissues, and hence facilitate assessing their status and state, is considerably augmented by its non-invasive character that allows enabling *in-vivo* assays[31, 32]. Among these past efforts, intravital and *in-vivo* imaging of the eye with SHG microscopy has also been successfully demonstrated in past works[33, 34], including for human patients [35].

Although the cornea it's composed of five layers, the stroma makes up approximately 90% of the corneal thickness [36] and it mainly consists on type I collagen fibers, presenting a particular organization to maintain the cornea's shape, transparency, and biomechanics[37]. SHG imaging of corneal collagen has been addressed in a significant number of works reported to date, which have demonstrated that various conditions of the cornea and of the eye can be identified, diagnosed and classified[7, 11, 12, 14, 33, 34, 38-40] with this emerging imaging tool based on non-linear optical effects. Some of the main advantages of this technique over other investigation modalities are: collagen-specific imaging, non-invasive optical sectioning (and consequent 3D imaging), and high-penetration depth enabled by the use of (pulsed) infrared lasers[15, 41, 42]. Past studies covering the topic of SHG imaging of the eye and cornea have demonstrated the usefulness of SHG imaging for better understanding the structural differences between healthy and edematous corneal tissues[40, 43, 44], while others have successfully explored associated conditions such as subepithelial fibrosis[45]. These past efforts demonstrate SHG as a highly promising tool to enable next-gen frameworks for *in-vivo* cornea characterization and diagnostics. Despite the availability of imaging systems fit for SHG investigations in clinical settings[46-48], the transformation of SHG imaging into a routine tool for clinical diagnostics faces near-future challenges, such as the difficulty of data interpretation. In addition to the moderate-to-high costs of fs-laser sources, an important bottleneck is the low availability of public SHG data sets, which have only recently started to emerge[49], alongside training resources that physicians would require to master diagnostic protocols enabled by this imaging technique. In efforts to alleviate such issues, a variety of automated methods for SHG data classification have been reported over the past years, building both on traditional machine-learning methods[50-56] and also on trending Deep Learning approaches, addressing the characterization of various parts of the human body such as ovarian tissues[57], prostate tissues[58], skin[18], breast[59] or liver[60]. However, to the best of our knowledge, classification of SHG images collected on eye tissues by machine or Deep Learning has not been explored to date.

Here we extend the current state-of-the-art on SHG imaging of the cornea by showing that SHG images collected on edematous and healthy cornea tissues can be efficiently classified with two highly popular pre-trained Deep Learning models known for their superb transfer learning capabilities[59, 61-65]: InceptionV3, ResNet50, and with a model that we custom-built, named Flexible Light-weight model for Bioimage Analysis (FLIMBA) that is trained from scratch. We introduce and evaluate different augmentation strategies tuned to the specifics of the application at hand and show that the three evaluated models provide complementary information in terms of image classification, as different image regions contribute to their decision-making. We discuss the stand-alone classification performance of InceptionV3, ResNet50 and FLIMBA, and show that their complementary nature can boost the overall classification performance up to 0.98 Area Under Receiver Characteristic (AU-ROC), by combining their outputs in a simple majority vote scheme. We consider this finding important, given its potential to stimulate the development of future Deep Learning methods for SHG imaging that make use of multiple-expert systems[66], which could compensate the current lack of SHG-specialized Deep Learning classification frameworks.

Finally, our approaches and results hold potential to be extrapolated to other diagnostics scenarios that urgently require new solutions, such as differentiation of corneal edema in different stages, automated extraction of hydration level of cornea, or automated identification of corneal edema causes. Our work can thus pave the way for novel methods for cornea diagnostics with Deep-Learning assisted Non-Linear Optical Imaging.

## 2. Methods

### 2.1 Image acquisition setup

A schematic diagram of the SHG imaging microscope used in this work is shown in **Fig. 1**. Detailed descriptions have been given previously[67]. Briefly, this instrument was built in-house, and relies on a Ti:Sapphire laser source (Mira900f; Coherent, St. Clara, CA) and a commercial inverted microscope (Nikon TE2000-U; Nikon, Tokyo, Japan). The 800 nm output of the laser was used for sample excitation, and the images were acquired using a non-immersion long working-distance objective (Nikon ELWD; numerical aperture [NA] = 0.5, 20×). The laser power that we used ranged between 50 and 100 mW at the sample plane (depending on the specimen under study) and the beam was scanned across the sample with an XY scanning unit (VM1000; GSI, Billerica, MA). SHG signals were acquired in the backward direction using the same objective. With this backward collection geometry, the signals passed through a dichroic mirror, and were filtered by band-pass (400±5 nm) filter. Finally, the signal reached the detection unit composed of a photomultiplier tube (PMT; R7205-01; Hamamatsu, Shizouka, Japan). The scanning time was set to approximately 1 s for a 200×200 pixels image. Image areas of 180×180 μm$^2$ were recorded.

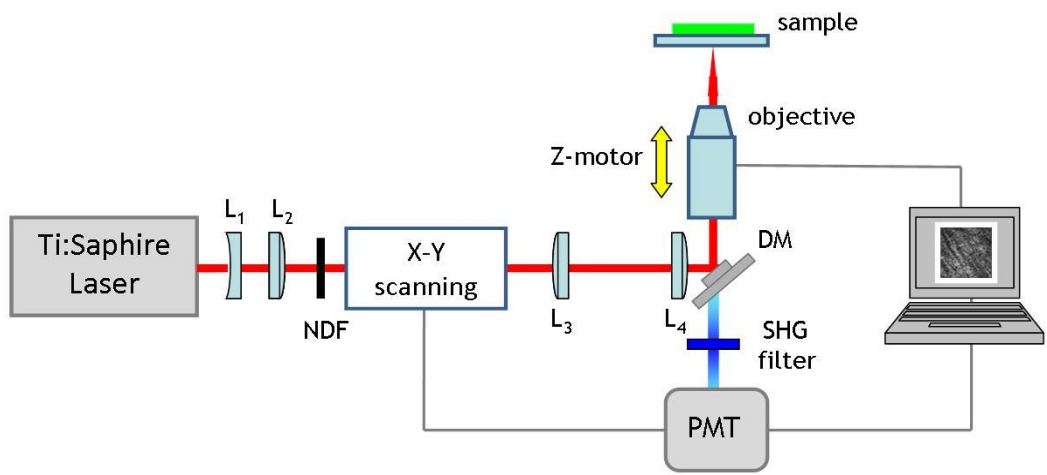

**Figure 1.** Sketch of the SHG microscope used for the purpose of this work. $L_1$-$L_4$, lenses; NDF, neutral density filter; DM, dichroic mirror; PMT, photo-multiplier tube. Further details about the different components and optical elements can be found in[67, 68].

### 2.2 Sample preparation

The porcine eyes used herein were obtained from a local slaughterhouse in Murcia, Spain, on the same morning of the experiment and moved to the laboratory immediately after death. Corneas were excised from the ocular globe with a trephine. Neither fixation, nor staining was used. These were placed upside down on a glass bottom dish (thickness: 170 μm) filled with phosphate buffer saline (PBS) solution and placed on the microscope stage for SHG imaging. Once fresh corneas were imaged, the specimens stayed immersed in PBS for several hours. During this time overhydration was produced, what allowed tissue swelling. Then the cornea became edematous and the collagen arrangement was modified, as discussed in more detail in previous works[40, 69].

**2.3 Data augmentation**

The dataset consists of 209 SHG images. The images are grouped in two classes: control and edema, which were well balanced, with 101 images from the control class and 108 from the edema class. The use of datasets of reduced dimension has been as well reported in past SHG Deep Learning methods[18, 57, 60], given the typical low data availability. As large datasets are known to favour high-performance for Deep Convolutional Neural Networks (CNNs) [70], image augmentation was used to synthetically increase the number of images in the dataset, so that the considered Deep Learning models generalize well after training. Specifically, we combined a widely common augmentation scheme based on 90-degree rotation that is known to help deep CNNs to generalize for affine transformations[71], with a Gaussian blurring scheme, using three different values for the standard deviation (σ). The latter strategy is meant to help the deep CNNs generalize for the size variability of the collagen structures which occur in different samples. The rationale for augmenting SHG datasets by Gaussian blurring has been introduced in our past work[18], and is discussed in more detail under Results. The employed augmentations are depicted in **Fig. 2**.

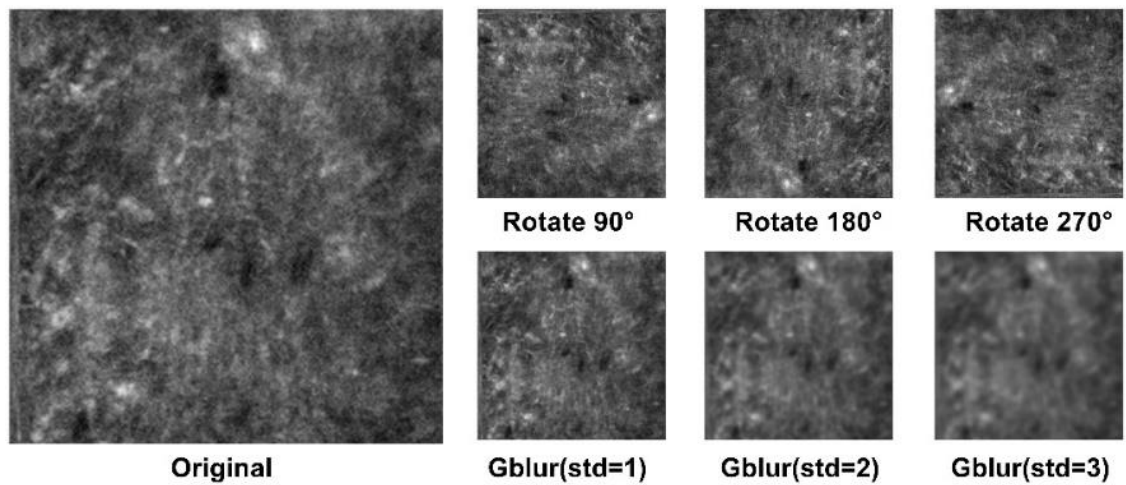

**Figure 2:** Augmentation of SHG data by rotation and Gaussian Blur. Outputs of "rotation" + "Gaussian blur" are not displayed.

Considering that Gaussian blurring does not represent a common augmentation method in Deep Learning classification tasks addressing natural images, to better assess and understand the effectiveness of combining this type of augmentation with rotation (a standard transformation used for data augmentation), we have constructed 11 datasets, each considering different rotation and Gaussian blur settings. These are presented in **Table 1**, together with the total number of images resulted in each dataset.

| Dataset name | Augmentations | Number of images |
|---|---|---|
| Dataset 1 (D1) | Rotation | 836 |
| Dataset 2 (D2) | Rotation and Gaussian blur (σ=1) on non-rotated images | 1045 |
| Dataset 3 (D3) | Rotation plus Gaussian blur (σ=1) on all images | 1672 |
| Dataset 4 (D4) | Rotation and Gaussian blur (σ=2) on non-rotated images | 1045 |
| Dataset 5 (D5) | Rotation plus Gaussian blur (σ=2) on all images | 1672 |
| Dataset 6 (D6) | Rotation plus Gaussian blur (σ=3) on non-rotated images | 1045 |
| Dataset 7 (D7) | Rotation plus Gaussian blur (σ=3) on non-rotated images | 1672 |
| Dataset 8 (D8) | Rotation plus Gaussian blur (σ=1) on all images plus another Gaussian blur (σ=2) on all images | 2508 |
| Dataset 9 (D9) | Rotation plus Gaussian blur (σ=1) on all images plus another Gaussian blur (σ=3) on all images | 2508 |
| Dataset 10 (D10) | Rotation plus Gaussian blur (σ=2) on all images plus another Gaussian blur (σ=3) on all images | 2508 |
| Dataset 11 (D11) | Rotation plus Gaussian blur (σ=1) on all images plus another Gaussian blur (σ=2) on all images plus another Gaussian blur (σ=3) on all images | 3344 |

**Table 1:** Augmentations for each data set and resulting number of images in the data set.

## 2.4 Deep Learning Models

Binary classification of SHG images collected on control and edematous porcine ocular tissues was experimented with three models. The first two models represent two widely used Deep CNNs: InceptionV3 and ResNet50, which are by default pre-trained with large datasets comprised of natural images and were fine-tuned in this experiment by training their final layer with the datasets presented in Table 1. InceptionV3 and ResNet50 were chosen due to their success in the ImageNet Large Scale Visual Recognition Challenges, and their proven efficiency for transfer learning [59, 61-65]. The third model that we used is a custom-developed one, that does not benefit of prior pre-training, and is trained from scratch using the Datasets assembled in this work. This model is coined Flexible Light-weight Model for Bioimage Analysis (FLIMBA) and was mainly developed in the purpose of experimenting the capacity of convolutional neural network to learn from small-scale datasets, which represent the usual premises in applications that rely on imaging techniques that are not widely available (due to their costs, lifetime, level of expertise and knowledge required, etc.), such as non-linear optical microscopy techniques, and which are typically confined to limited number of developing groups. For all three models tested, the datasets presented in Table 1 were randomly split in two partitions, summing up 60% of all images for training and 40% for testing. A model was then trained on the training set and evaluated on the testing set. The model, train and test datasets were then deleted. We define this process as a training session. It is important to note that although the initial dataset for each session was the same, the training and testing set always differed because their entries were randomly selected every time. This way we can make sure that we train and evaluate the models on different distributions of the same dataset. Each training session was repeated 10 times. In the following we provide a series of details on the three models used.

*a) InceptionV3 & ResNet50*

The first version of the Inception networks (initially named GoogLeNet) was first introduced in 2014 by a team of researchers at Google [72]. The main advantage of this class of models is their ability to allow for parallel convolutions of different sizes while keeping the number of parameters relatively low (under 25 million). In the same year it was introduced it won the ImageNet challenge. The Inception class of models is often used in medical applications with weights pre-trained from ImageNet [18] [73]. The ResNet was first introduced in 2015 by a group of researchers at Microsoft [74] to solve the problem of vanishing gradients by introducing identity connections from the input to the end of each residual block. In the same year, it won the ImageNet challenge. It is also used in medical applications with weights pre-trained from ImageNet.[75]

In this work, we use the InceptionV3 and ResNet50 pre-trained on ImageNet for transfer learning. Both models were fine-tuned to cornea SHG imaging and binary classification by replacing the final layer with a SoftMax layer with 2 neurons, to accommodate for binary output, and all weights were unfrozen. During training, we use the Adam optimizer with a learning rate of 1e-5. Early stopping was used to prevent overfitting. All images that we used as input for InceptionV3 had to be upscaled to 299x299 pixels, and the pixel values had to be scaled between -1 and 1, to match the requirements of InceptionV3. To address the specifics of ResNet50, the images were upscaled to 244×244 pixels, converted from RGB to BGR, and each color channel was zero-centred with respect to the ImageNet dataset.

*b) Flexible Light-weight Model for Bioimage Analysis (FLIMBA)*

The FLIMBA model is inspired from recent work focusing as well on the development of artificial intelligence methods for classification of bioimages under the premises of small-scale training set availability[76]. In the design and development of FLIMBA we had in mind the small size of the dataset, avoiding overfitting, and obtaining a good generalization for different types of samples. All these can be solved by using dropout. It is well known that adding dropout layers prevents a neural network from overfitting[77], the advantages of using this strategy being more noticeable on small datasets given that all types of neural networks are prone to overfitting on small datasets[77]. The generalization problem was partially addressed by applying a multitude of augmentations on the original dataset, which is complementary to model level generalization that can be achieved by using dropout [78].

Considering the need and utility of dropout layers we composed the FLIMBA model as a seven-layer CNN, followed by a Global Average Pooling (GAP) layer, two fully connected (FC) layers, and a two neuron softmax layers. The model architecture is schematically displayed in **Fig. 3**. For training the FLIMBA model, we used L2 regularization with λ of 1e-4 and Adam optimizer with a learning rate of 1e-5. The pixels of all images used as input for the FLIMBA model were scaled between 0 and 1. Compared to ResNet50 and InceptionV3, the images used as input for the FLIMBA model were kept at their original resolution of 200×200 pixels.

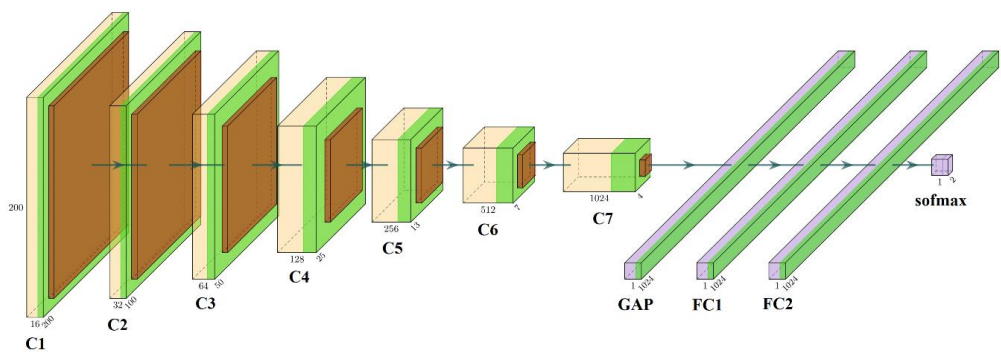

**Figure 3:** Schematic representation of the FLIMBA model. Each convolutional layer (light yellow) is followed by a dropout layer with 0.2 probability (light green), followed by a 2×2 max-pooling layer. After the last max-pooling layer, a GAP and two FC layers (light purple), are added, each followed by a dropout layer with 0.5 probability (green). Last, an output layer with two softmax neurons is used.

**c) Majority vote model**

Besides evaluating the performance of InceptionV3, ResNet50 and FLIMBA as stand-alone solutions for the classification of control and edematous corneas, we also investigated the combined use of the three models, given their complementarity (discussed under Results). To this end, we use each model to make individual classifications and then use a basic majority vote scheme to decide the output of the Majority vote classification strategy.

**3. Results and discussions**

**3.1 Qualitative assessment**

The corneal stroma is composed primarily of type-I collagen[79, 80], which is generally acknowledged as a strong source of SHG signals. This is confirmed by the backscattered SHG signals that can be observed in **Fig. 4**, which illustrates representative examples of the SHG images collected in this experiment.

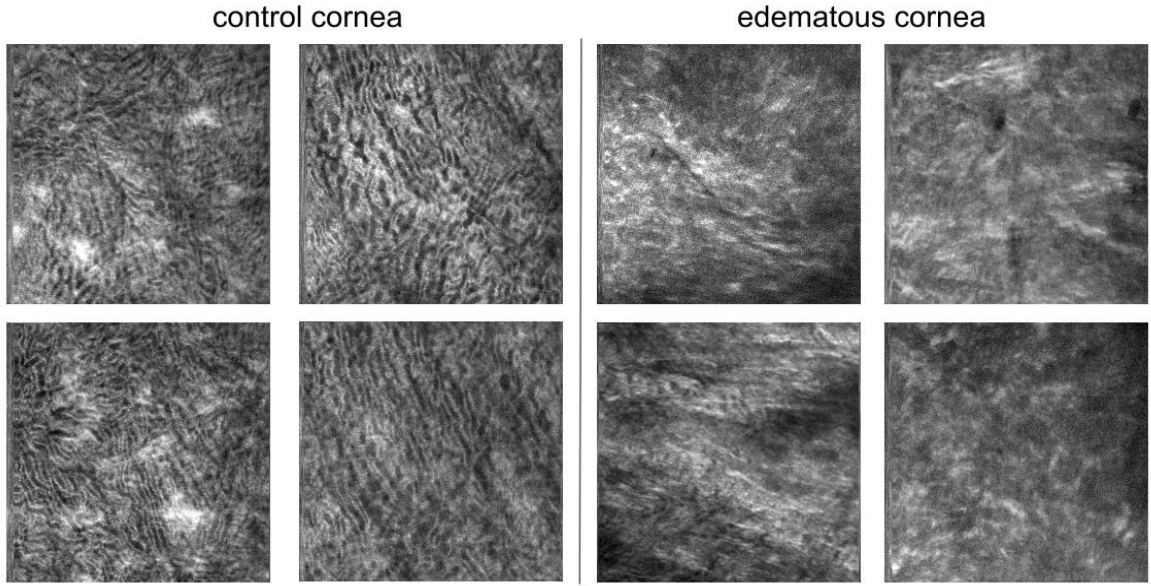

**Figure 4:** SHG images collected on control and edematous porcine eye cornea. Field of view: 180×180 μm$^2$.

In agreement with past investigations using cornea SHG imaging[11, 40, 69], in the control sample we can observe narrow bundles of fairly linear structures corresponding to stromal fibers. It should be noted that previous studies have shown that the backward-SHG images (BSHG) of the normal cornea are morphologically different from the forward-SHG images (FSHG), whereas for the edematous cornea BSHG and FSHG images have a rather similar aspect[40]. In our experiment we investigate only the former imaging mode, backward-SHG, given that this mode can also be used for intravital / *in-vivo* imaging of the eye, and subsequent diagnostics, as shown in[33-35], whereas forward SHG is practical only for *ex-vivo* / *in-vitro* applications. In the case of the edematous cornea, a series of structural modifications resulting from the overhydration occur, such as increases in lamellae spacing, collagen denaturation, decrease in fiber delineation and loss of organization. These modifications are nonuniform throughout the stroma, depending on variations in the hydration. The images that we collected on edematous cornea reflect this situation, and we can observe that the narrow bundles observed in the control cornea images are no longer clearly visible, and the overall aspect of the image becomes non-uniform; the endogenous harmonophores are still present, given the SHG signal levels, but become disorganized. A more thorough interpretation of SHG images, such as those presented in **Fig.** 4, or in previous works [11, 40, 69], requires extensive collective knowledge on: i) cornea physiology;  ii) collagen architecture in healthy and edematous cornea and iii) the specifics of SHG imaging. This can represent an important issue for the further penetration of SHG corneal imaging into the clinical realm, where SHG still represents at the time being an "exotic" technique. In the next, we contribute to this problem by showing that SHG images of the cornea can be classified between the control and edema groups automatically, in a fully unsupervised manner, by modern artificial intelligence models.

## 3.2 Classification of control vs. edematous cornea with Deep Learning

### *3.2.1 On the complementarity of the tested models*

Over the past decade Deep Learning[81] methods have taken the realm of image analysis by storm, with new models reported on a frequent basis. Undoubtedly, the vast majority of models are developed taking into account the specifics of natural images, with most of their applications targeting tasks such as scene or object classification, identification of gestures, stances, etc. Many of these models have been also used for bioimage classification problems, as their transfer learning capabilities allow making use of knowledge learnt from natural images to classify bioimages collected with benchtop equipment used for lab research purposes, or with imaging instrumentation fit for clinical use. In such scenarios, a model that is pretrained with large-scale natural image sets, is fine-tuned by retraining its final layer(s) by using images specific to the application at hand (E.g. bioimages collected with a specialized instrument). A vast number of such efforts have been reported to date and have shown that for a particular problem some models work better than others. A thorough understanding of the performance variation is still missing, and discussions on the complementarity of distinct models are scarce in the literature, with respect to bioimage classification topics, and even more so for SHG applications.

In this work we evaluate three models, InceptionV3, ResNet50 and FLIMBA, which are significantly different in architecture and complexity. Besides our objective on assessing the performance of these models, we also wanted to understand if they provide complementary outputs, which can eventually be merged to enhance the overall classification performance. To this end we used a technique Gradient Class Activation Map (GradCam)[82] which provides a spatial map (heatmap) showing how intensely the input image activates the model, thus highlighting what image regions contribute most in the decision-making of the used model. The layout of the GradCAM method[82], can be briefly summarized as follows: 1) Compute the model output and last convolutional layer output for the image. 2) Find the index of the winning class in the model output. 3) Compute the gradient of the winning class with respect to the last convolutional layer. 4) Average this, then weigh it with the last convolutional layer. 5) Normalize between 0 and 1 for visualization. In **Fig. 5** we present the heatmaps computed as above described for a control and edema images. Similar results have been obtained for all images in the set.

In the heatmaps presented in **Fig. 5**, we can observe a significant difference in the spatial layout of the image regions that contribute most to the decision-making of a specific model. Such differences can have strong implications with respect to the final output, in the sense that a given image can be correctly classified by one model and incorrectly classified by another. Such situations can be made useful by employing multiple expert decision strategies[66], where a final decision for a classification task is made based on the output of several classifiers, by merging their individual decision by various methods such as majority vote. Considering the heatmaps that we obtained for the tested image set, which suggest strong complementarity of the three tested models, we were motivated to assess a majority vote scheme, where the class of an image is given by 2 out of 3 votes of the InceptionV3, ResNet50 and FLIMBA models. The obtained results are presented next.

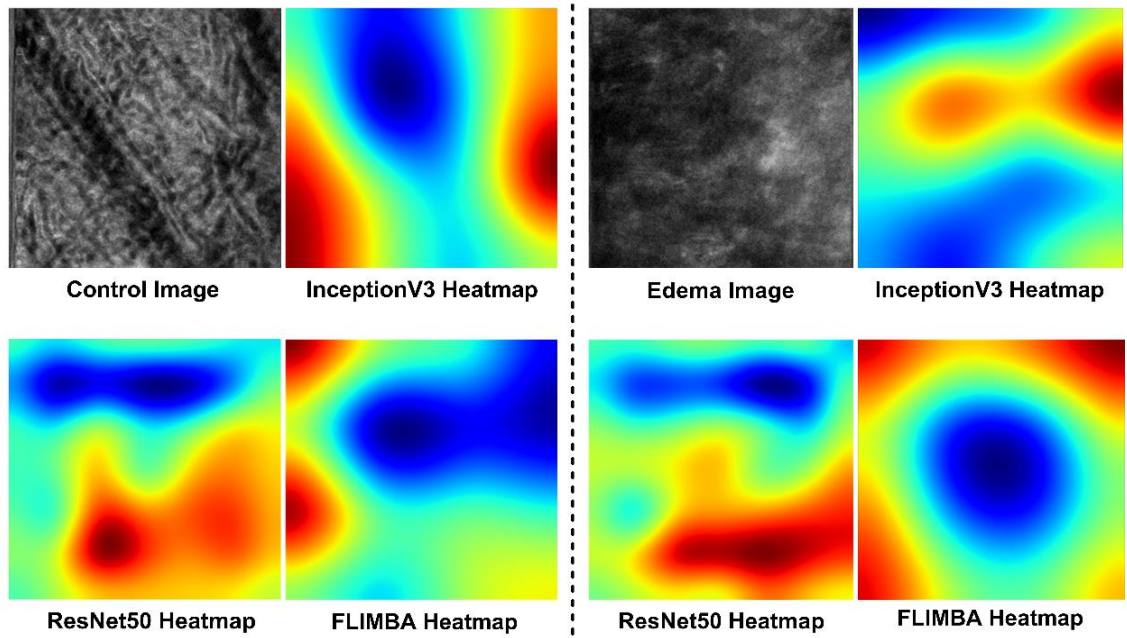

**Figure 5.** Activation heatmaps for the InceptionV3, ResNet50 and FLIMBA models calculated by GradCAM. The colors in the heatmaps indicate what image region contributes most in the model's decision-making process, with red regions representing regions of high importance in activation of neurons whereas blue indicate image regions of least importance.

*3.2.2 Classification accuracy under different data augmentation strategies*

Here we present the results obtained using the InceptionV3, ResNet50 and FLIMBA models, together with those obtained using a majority vote scheme where 2 out of 3 votes of the considered models are used for assigning a class. All 11 datasets, built under different augmentation strategies (Table 1), have been tested and the results are presented in **Fig. 6**. The best performing data set is D11, with ResNet50, FLIMBA and InceptionV3, providing 97.42 ±1.98%, 93.75 ±3.26% and 91.78 ±1.58%, respectively, average accuracies for 10 runs, and the Majority vote scheme providing an average accuracy of 98.14 ± 0.52%.

Based on the results presented in **Fig. 6** we observe the following:

-an increase in the size of training data set corresponds to an increase in the achieved accuracy. This behavior is most noticeable when looking at the accuracies obtained for the D1 (863 images) and D11 (3344 images) datasets.

-As the standard deviation ($\sigma$) of the applied Gaussian blur increases, the accuracy decreases. D2, D4, and D6 have the same number of images but in these three cases the Gaussian blur augmentation is performed with increasing $\sigma$. The same can be observed for D3, D5, D7.

-When augmenting with multiple Gaussian blur kernels, using kernels of lower σ provides better results compared to using higher σ kernels. See accuracies of D8, D9 and D10.

-Although Gaussian blurring with higher σ results in worse classification performance compared to using lower σ, datasets generated with lower σ are nonetheless useful as they hold additional, complementary, information. This can be best observed for Datasets 10 and 11, where the additional Gaussian blurring performed with a kernel of σ=3, results in better accuracy compared to using only σ=1 and σ=2 kernels.

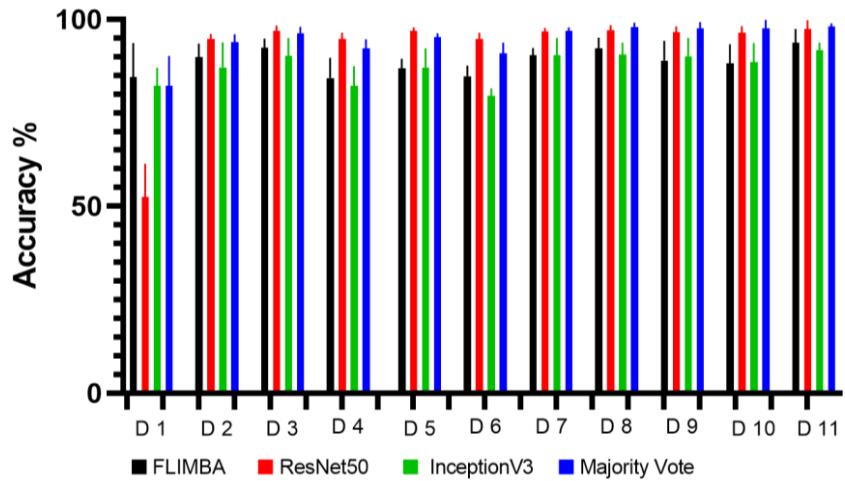

**Figure 6:** Accuracy for InceptionV3, ResNet50, FLIMBA, and Majority Vote on all 11 Datasets.

Concerning the relevance of Gaussian blurring for microscopy images collected on tissues, we refer to our previous work[18], where we explained the advantages provided by such augmentation strategies by referring to the fact that different layers of the Gaussian image pyramid approximate images of the original objects collected at different scales[83]. Thus, by including layers of the image pyramid in the training process approximates exposing the model to structures of different sizes, conferring thus scale invariance to the classification framework. This is particularly important to achieve given the variation in size of anatomical structures, depending on age, weight, or anamnesis of the patients[84-86] and other factors. Convolving laser microscopy images with a Gaussian filter is also known to suppress noise[87], which might as well contribute to an increase in the classification performance. Overall, the here presented results show the importance of Gaussian blurring as an augmentation strategy for microscopy bioimage datasets and shed additional light on using this augmentation strategy in an efficient manner to assist bioimage classification with Deep Learning.

*3.2.3 Additional insights on the classification performance using InceptionV3, ResNet50 and FLIMBA*

As presented in **Fig. 6**, best accuracy was observed for Dataset 11 comprised of 3344 images, summing up original images, rotated original images, and Gaussian blurred (σ=1, σ=2, σ=3) instances of all

(original + rotated) images. In **Fig. 7** we provide additional details on the classification performance of the three considered models, and their majority vote for this Dataset. As previously discussed, among the three tested models ResNet50 provides best accuracy (97.42±1.98%), followed by FLIMBA (93.75±3.26%), and then InceptionV3 (91.78±1.58%); the Majority vote scheme yields higher accuracy than all three tested models (98.14±0.52%). Similar Accuracy trends could be observed for the other tested Datasets, **Fig. 6**. ResNet50 also excels in Sensitivity (0.993±0.006), followed by InceptionV3 (0.989±0.008), and FLIMBA (0.916±0.060). The majority vote performs this time worse (0.984±0.005) than InceptionV3 and ResNet50, due to the poor Sensitivity performance of FLIMBA, which biases the performance of this decision-making strategy. ResNet50 provides as well best results (0.959±0.031) in terms of Specificity, followed by FLIMBA (0.954±0.042) and Inception V3 (0.824±0.102). The majority vote provides better Specificity (0.965±0.046) than any of the three individual models.

To provide a general overview incorporating all three metrics, while also considering different classification thresholds we use Receiver-Operator-Characteristic Curves (ROC), which represent a widely used metric for evaluating classification performance[88, 89]. The calculated AU-ROC values show best classification performance for ResNet50 (AU-ROC=0.975±0.018), followed by FLIMBA (AU-ROC=0.935±0.0518) and InceptionV3 (AU-ROC=0.918±0.055). The Majority Vote scheme yields a better classification performance than any of the three tested models, AU-ROC=0.983±0.025. Although the difference between the stand-alone use of ResNet50 and the Majority vote scheme is not high (AU-ROCs of 0.97 and 0.98, respectively), it should be noted that the difference of 0.01 AU-ROC represents 33% of the missing 0.03 in ResNet50's performance, thus we argue that the Majority vote represents an useful strategy to address classification gaps in the stand-alone use of a specific classifier, in the final goal to achieve classification performance as possibly close to the ideal AU-ROC= 1.

Based on the results depicted in **Fig. 7**, several aspects are interesting to observe. Although ResNet50 and InceptionV3 are both pre-trained on the ImageNet set, ResNet50 provides considerable better classification performance, indicating more efficient transfer learning capabilities for the here considered application. While FLIMBA benefits of no apriori training, being trained from scratch, it provides satisfactory classification results, suggesting that the development of simple models, tuned to the specifics of the SHG bioimaging application at hand, represents a subject worthy the be explored in more detail. Finally, despite the simplicity of the majority vote scheme that we used, it provided better overall results than any of three tested models. We consider this finding to be important as it suggest that the complementarity of Deep Learning models built on different architectures and concepts makes their joint exploitation useful, irrespective of their performance variability. These results indicate the potential usefulness of more elaborate multiple-expert decision strategies for AI assisted SHG data classification. These results suggest as well that Ensemble Learning[90, 91] strategies could prove very efficient in classification of SHG bioimages.

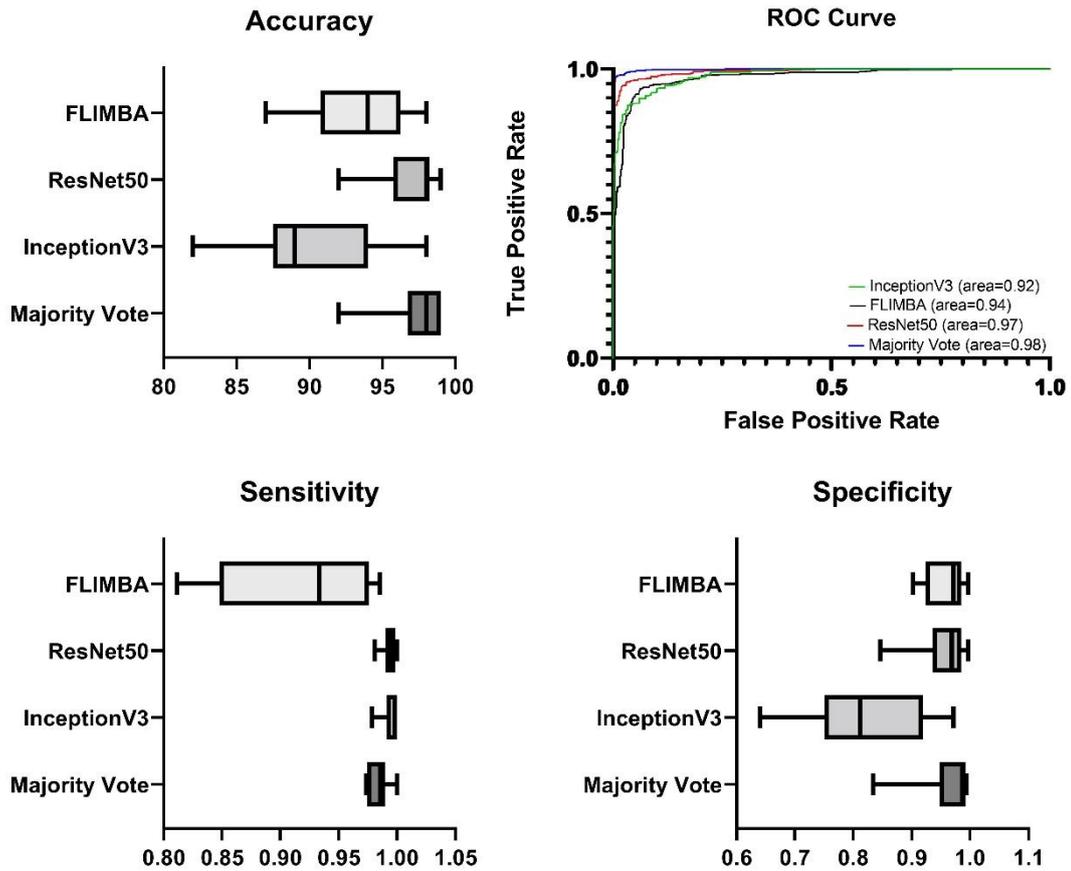

**Figure 7:** Box and whisker plot for the accuracy, sensitivity, specificity, and ROC curve for InceptionV3, ResNet50, FLIMBA, and Majority Vote for Dataset 11. Accuracy, Sensitivity and Specificity are calculated at the default binary classification problem threshold of 0.5. (The whiskers indicate smallest and largest values; the box extends form the 25$^{th}$ to 75$^{th}$ percentiles and the vertical line is drawn at the median).

## 4. Conclusions

Overhydration of the cornea above its 78% physiological level, an affection known as corneal edema, can result in significant discomfort for the patient, followed by important complications, if not addressed timely. This condition is typically associated to compromised anatomic and/or functional integrity of the corneal stroma. Given the cornea's high collagen content, SHG imaging has been demonstrated in past works as a very valuable tool to image and characterize the pathological status of corneal tissues, with potential to be used in both research studies aimed at providing a better understanding of this part of the eye, and in diagnostics. With respect to the latter, SHG data interpretation difficulties, still pose significant problems for the translation of SHG to the clinical practice. In this work, we have demonstrated for the first-time classification of edematous corneal tissues with Deep Learning assisted SHG. We explored the use of three distinct, and highly different models, ResNet50, InceptionV3 and FLIMBA (a custom developed model), and discussed their complementarity and classification performance, next to data augmentation strategies addressing the specifics of the application at hand. Importantly, we showed that the combined use of these three models in a simple Majority Vote Scheme provides better results, compared to the stand-alone use of any of these. This result indicates that multiple-expert decision strategies represent an important route to follow in the future in the context of SHG bioimage classification. We consider this work to provide important insights on the use of Deep Learning models for the automated classification of SHG images, and of non-linear optical images, in general. Besides inspiring other related application in SHG imaging, we expect that our results will facilitate and stimulate future developments considered highly important in ophthalmology such as classification of corneal edema in different stages, automated extraction of hydration level of cornea, or automated identification of corneal edema causes, representing topics that are in search of urgent solutions.

**Acknowledgment:** SGS, RH and GAS acknowledge the support of UEFISCDI Grant PN-III-P2-2.1-PED-2019-1666. MJH acknowledges support from the Flagship of Photonics Research and Innovation (PREIN) funded by the Academy of Finland (Grant No. 320165). JMB acknowledges the supports from Agencia Estatal de Investigación, Spain (grant PID2020-113919RB-I00).